\documentclass[%
 reprint, superscriptaddress,
 amsmath,amssymb,
 aps,
]{revtex4-2}

\usepackage{graphicx}
\usepackage{dcolumn}
\usepackage{bm}
\usepackage{xcolor}

\begin{document}
\title{\textbf{Probing and modeling cell-cell communication in 2D biomimetic tissues}}

\author{C. M. Vincent}
\address{Laboratoire Jean Perrin, UMR 8237 Sorbonne Université/CNRS, Institut de Biologie Paris Seine, 4 Place Jussieu, F-75005 Paris, France}
\author{S. Ravindran}%
\address{Laboratoire Jean Perrin, UMR 8237 Sorbonne Université/CNRS, Institut de Biologie Paris Seine, 4 Place Jussieu, F-75005 Paris, France}
\address{Laboratoire Matière et Systèmes Complexes, CNRS UMR 7057, Université Paris Cité, 10 Rue Alice Domon et Léonie Duquet, 75013 Paris, France}
\author{A. M. Prevost}
\address{Laboratoire Jean Perrin, UMR 8237 Sorbonne Université/CNRS, Institut de Biologie Paris Seine, 4 Place Jussieu, F-75005 Paris, France}
\author{L.-L. Pontani}
\address{Laboratoire Jean Perrin, UMR 8237 Sorbonne Université/CNRS, Institut de Biologie Paris Seine, 4 Place Jussieu, F-75005 Paris, France}
\author{O. Bénichou}
\address{Laboratoire de Physique Théorique de la Matière Condensée, UMR 7600 Sorbonne Université/CNRS, 4 Place Jussieu, F-75005 Paris, France}
\author{E. Wandersman}
\email{elie.wandersman@sorbonne-universite.fr}
\address{Laboratoire Jean Perrin, UMR 8237 Sorbonne Université/CNRS, Institut de Biologie Paris Seine, 4 Place Jussieu, F-75005 Paris, France}

\date{\today}

\begin{abstract}
In tissues, cells in direct physical contact with each other can exchange ions or molecules via protein clusters called gap junctions that form channels across the membranes of adjacent cells. Here, we use a simplified biomimetic approach, coupled with theoretical modeling, to unravel the physical mechanisms controlling such transport. Tissues are mimicked with 2D hexagonal networks of monodisperse aqueous droplets connected by lipid membranes called Droplet Interface Bilayers (DIBs), decorated with $\alpha$-Hemolysin ($\alpha$HL) transmembrane proteins forming nanopores through heptamerization in the membrane. The diffusion of calcein across 2D DIB networks is thoroughly studied using epifluorescence microscopy at various $\alpha$HL concentrations. The results are successfully confronted with a Continuous Time Random Walk model in hexagonal networks, with an average waiting time increasing nonlinearly with the concentration of pore monomers.
\end{abstract}

\maketitle

\section{Introduction}
All cells can perceive information on the chemical, electrical and mechanical properties of their immediate environment~\cite{phillips2012physical,potter2016communication}. This information plays a crucial role in regulating essential cellular functions, such as growth, division, apoptosis, and motility. Although isolated cell populations subjected to a chemical field often exhibit stochastic and heterogeneous responses, their behavior becomes markedly more coordinated within a tissue~\cite{sun2012spatial}. Through direct cell–cell contact and communication, populations of cells can synchronize their dynamics and generate faster, cooperative responses to external cues. Another example is morphogenesis, during which intercellular communication mediated by the transport of morphogens, coordinates division, growth, and differentiation across multiple cells, ultimately guiding the formation of a functional organ~\cite{ellison2016cell,david2024formation,sahu2021germline}.
\\

Cell-cell communication encompasses various mechanisms by which cells exchange information. Such communication occurs via different modes depending on whether cells are physically connected or not. When they are separated, the communication can be mediated by ion channels, which are protein nanopores embedded in their plasma membrane. In this case, the transport of ions and molecules occurs by diffusion in the extracellular medium from cell to cell~\cite{sun2012spatial}. Communication between separated cells can also occur via physical links, such as tunneling nanotubes joining different cells~\cite{zurzolo2021tunneling,ariazi2017tunneling} which coordinate metabolism and signaling. Conversely, when cell membranes are in direct physical contact (like in epithelial tissues), communication is ensured by clusters of proteins creating a nanometric channel called ``gap junction”  that directly connect the interiors of two adjacent cells \cite{ariazi2017tunneling,schultz1986membrane,nielsen2012gap}. Discovered in the 1960's by Loewenstein and colleagues\cite{loewenstein1966permeability}, these structures were found to maintain the electrical continuity of tissues and allow the diffusion of molecules, proteins and even RNA~\cite{wolvetang2007gap} across the tissue. Gap junctions are found in various types of tissues, including epithelial layers, cardiac muscle, and smooth muscle, where they facilitate direct intercellular communication \cite{Bruzzone1996,Nielsen2012,Kanaporis2020}.\\
Fully characterizing the transport of ions and molecules within cell populations is complex, as it arises from multiple, often intertwined communication modalities~\cite{ariazi2017tunneling}. A central challenge is therefore to disentangle this communication network and isolate the contribution of specific signals. In addition, even for a sole type of communication mode, the physical laws controlling ion and molecular transport  can also be complex themselves : in gap junctions, the transport of molecules from cell to cell depends on the type of connexin, the size and charge of the permeant molecules, as well as the pH or ions concentration. It also depends on the maturation state of the junctions after their formation and their concentration which is itself regulated by the cells~\cite{neyton1986physiological,otero2016symplastic}. But even from a physical, reductionist point of view, the mechanisms that set the transport capacity across a given junction have not been fully described and quantified at the microscopic scale, nor the dependence of the transport properties with the concentration of gap junctions in the cell membrane. Moreover, at the tissue scale, one may wonder how the diversity in shape and topology of the tissue does influence the transport properties at mesoscopic length scales. Therefore, quantitative measurements, coupled to a statistical physics approach to describe diffusion across gap junctions in tissues remain to be implemented. \\
In this paper, we mimic the gap junction mediated cell-cell communication in tissues using 2D hexagonal lattices of aqueous droplets connected by lipid membranes decorated with nanometric channels. We measure quantitatively the diffusion of fluorescent molecular probes in the network and propose a random walk model to interpret the data, aiming at bringing a physical framework for molecular diffusion across gap junctions in tissues. Fitting our diffusion data with this model, we extract a characteristic diffusion time and we find that it increases non linearly with the concentration of nanopores.

\section{Materials and methods}
\subsection{Aqueous and oil/lipid phases}

Artificial tissues are made from Droplet Interface Bilayer (DIB) networks \cite{bayley2008droplet,villar2013tissue,valet2019diffusion}. Briefly, they consist of aqueous droplets (size $2R\sim$ 160 $\mu$m) bathing in an oil and lipids mixture. Lipids first form a monolayer at the oil/droplet interface (on timescales of the order or minutes). If two droplets are put in contact, a planar lipid bilayer spontaneously forms at their interface in which transmembrane protein nanopores can be inserted. \\
The aqueous phase consists in a 100 mM KCl solution buffered with 10 mM HEPES at pH 7.4 (unless stated, all chemicals were purchased from Merck inc.). Alpha-Hemolysin ($\alpha$HL) monomers are dissolved in the aqueous phase (mass concentration from 50 to 150 $\mu$g/mL).  For fluorescent aqueous droplets we add Calcein at 0.2 mM in final solution, which falls in a range of concentration in which the fluorescence intensity grows linearly with the calcein concentration \cite{hamann2002measurement}.\\
The oil phase is a 50/50 v/v mixture of Hexadecane and AR20 Silicon oil in which a lipid mix is dissolved. The lipid mix is a mixture of 4 different lipids, as in \cite{dupin2019signalling}. It is first prepared in liquid chloroform in a vial, to reach a final concentration of 4 mM DOPC (1,2-dioleoyl-sn-glycero-3-phosphocholine), 0.5mM DPhPC (1,2-diphytanoyl-sn-glycero-3-phosphocholine), 0.25mM DOPG (1,2-dioleoyl-sn-glycero-3-phosphoglycerol) and 0.25mM cholesterol. The mixture is aliquoted in vials containing 6.5 mg of total lipid weight. The chloroform lipid mix is then dried under gentle nitrogen flow. The dried lipid film covers the vial walls and the vial can be stored in a -20\textdegree C freezer for months. To obtain the final oil+lipids phase, 1 mL of the Hexadecane/Silicon oil solution is added to a vial  (6.5 mg/mL final lipid concentration) and the solution is placed in an ultrasonic bath for 30 minutes at a temperature of 30\textdegree C. The oil+lipid solution is kept for up to 3 days and vortexed immediately before use.\\ 

\subsection{Droplet Printing and network formation}
\begin{figure*}[htbp]
\centering
  \includegraphics[width=0.8\textwidth]{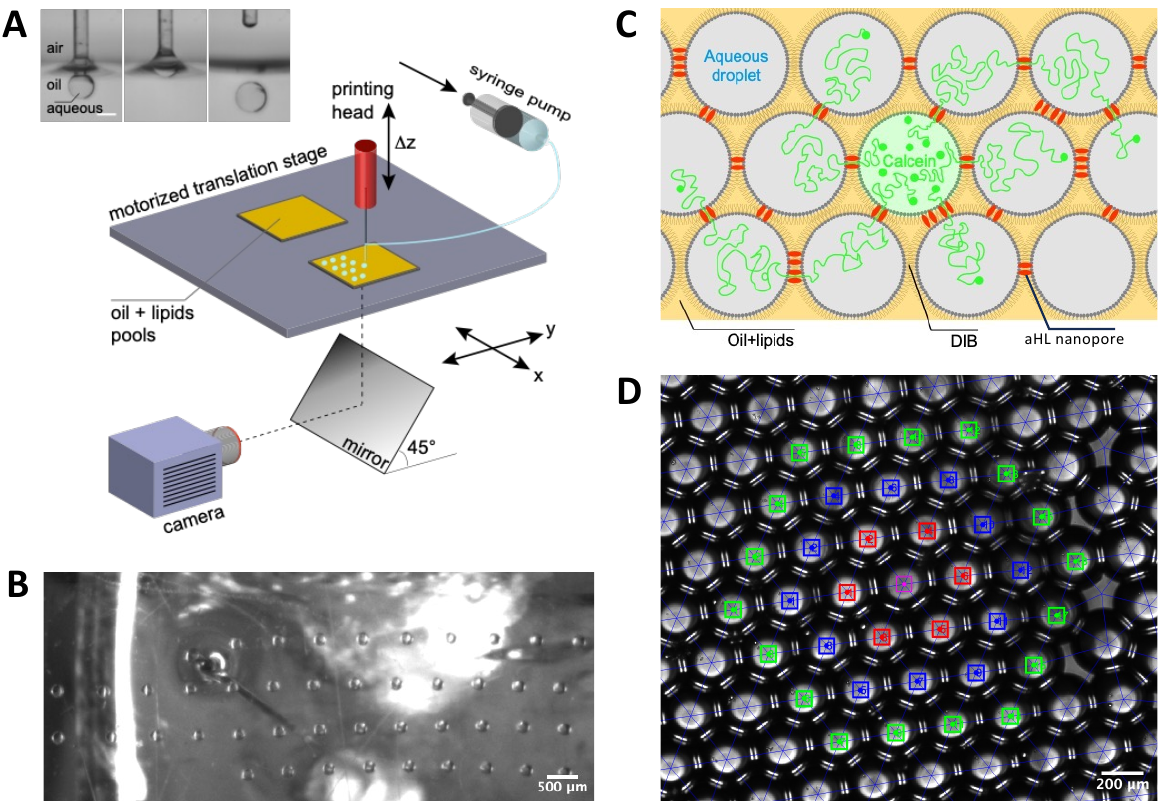}
  \caption{\textbf{(A)} Sketch of the droplet printer setup. The printing head, mounted on a vibration exciter is positioned above an oil+lipid pool placed on the XY motorized translation stage.  A syringe pump imposes a flow of the aqueous phase through a capillary tubing, attached to the printing head. Inset: principle of the printing technique -- images reproduced from \cite{valet2018quasistatic} showing the detachement of an aqueous droplet as it crosses an oil/air interface (the white scale bar  is 400 $\mu$m long).  \textbf{(B)} Droplets printed using the setup imaged from below. \textbf{(C)} Principle of the experiment: the Calcein molecules contained in a source droplet diffuse in the network of DIBs connected with $\alpha$Hemolysin nanopores. \textbf{(D)} The DIB network after compacting the droplets observed in bright field. The source (magenta symbol), first (red), second (blue) and third (green) neighbors positions have been overplotted on the image. The thin blue lines represent the Delaunay tessellation used to determine the connectivity of the droplets.}
  \label{Fig1}
\end{figure*}

The DIB network is made using a homemade droplet printer (Fig.~\ref{Fig1}a) in a Plexiglas square pool (dimension 10x10 mm) containing 340 $\mu$L of the oil phase. The printing process relies on the capillary trap instability that we have discovered and fully described in~\cite{valet2018quasistatic,valet2019diffusion}. Briefly, it consists in injecting at constant flow rate ($Q=25 \mu$L/h) the aqueous phase through a thin glass capillary (Molex, Polymicro Technologies capillary tubing, inner diameter 40 $\mu$m) connected to a flexible tubing, using a syringe pump (KDS Scientific). The capillary tip is attached to a Z vibration exciter (Brüel \& Kjaer, type 4810) allowing to oscillate the tip at $f$=2 Hz across the oil/air interface and a typical amplitude $\approx$ 2 mm. Briefly, when the tip is immersed in the oil phase, an aqueous droplet grows, and when the tip crosses back the oil/air interface, the droplet detaches and sink in the oil phase due to surface tension contrasts of the three different phases~\cite{valet2018quasistatic}. The pool is positioned on a XY motorized translation stage (Tango Desktop Märzhäuser) to move the container at each oscillation cycle, in order to produce well separated droplets (\textit{see} Fig.~\ref{Fig1}b and Supplementary Movie S1). Using this method, we print typically a hundred droplets with a mean diameter 2$R$=157 $\pm$ 10 $\mu$m (Fig.~2). The separated aqueous droplets are let for a few minutes to allow for lipid interface maturation  and the pool is then gently tilted by a few degrees to pack by gravity the droplets together and form the DIB network (\textit{see} Fig.~\ref{Fig1}c and Supplementary Movie S2). Once the lipid DIB membranes are formed, the $\alpha HL$ monomers present in all the droplets can heptamerize in the membrane to form nanopores\cite{valet2019diffusion}. In practice, we repeat the whole process (printing and tilting) twice to obtain wide enough ($\sim$ 1 cm) networks. In the second repetition, a few (typically three) widely separated fluorescent droplets containing Calcein are added. The final DIB network consists in these isolated fluorescent source droplets embedded in a network of non-fluorescent ones.\\  

\begin{figure}[htbp]
\centering
  \includegraphics[width=0.45\textwidth]{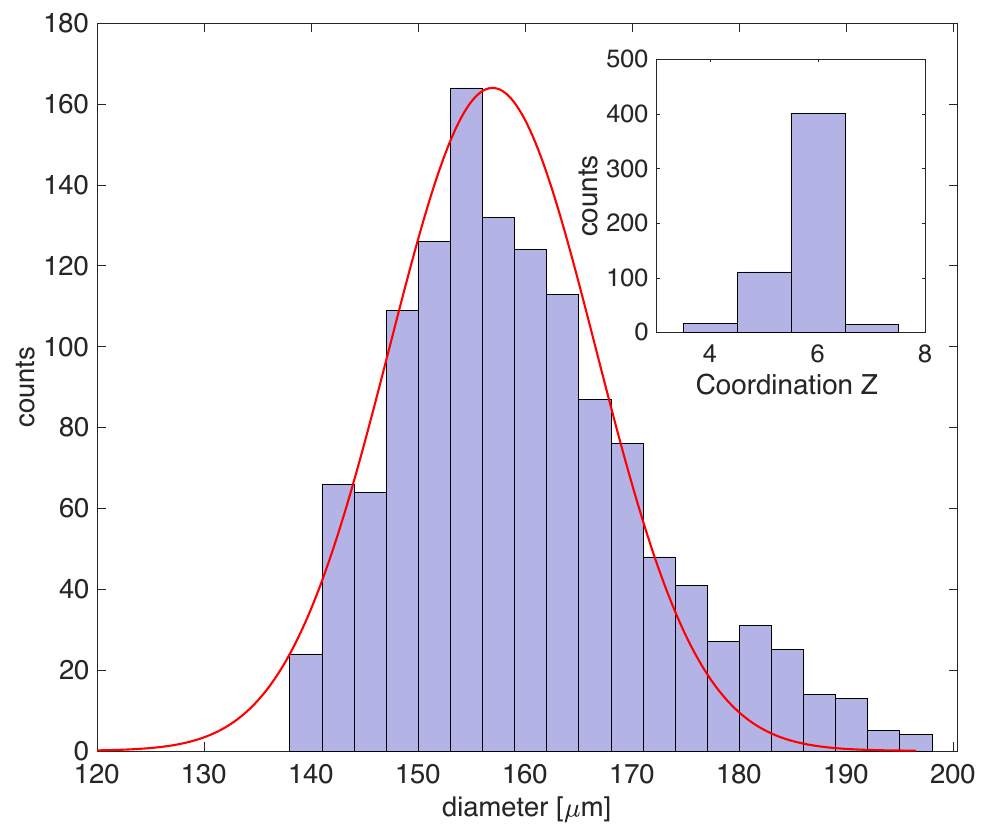}
  \caption{Histogram of the droplet diameters. The solid line is a gaussian fit, yielding $\bar{d}$ = 157 $\pm$ 10 $\mu$m. Inset: histogram of the coordination number.}
  \label{Fig1B}
\end{figure}

\subsection{Imaging}
\begin{figure}[htbp]
  \includegraphics[width=0.5\textwidth]{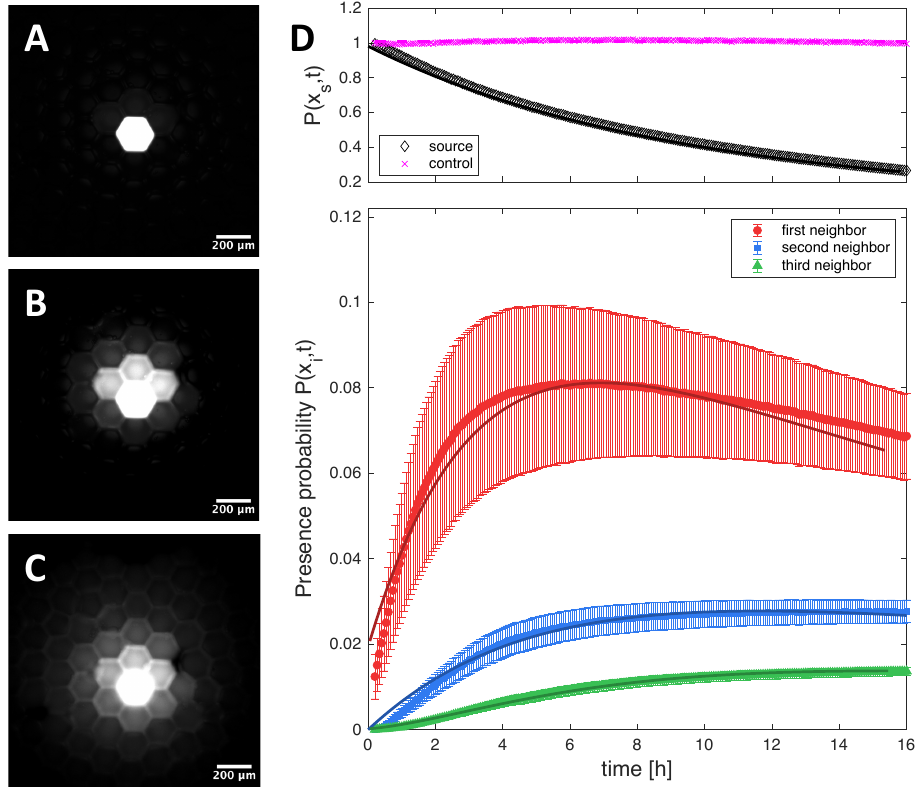}
  \caption{\textbf{A-C} - Fluorescence images of a DIB network, with $c_m$=125$\mu$g/mL at different times ($t$=0; 6 and 16 hours, respectively) showing the diffusion of calcein in the network. D) The corresponding normalized intensity (presence probability) as a function of time, for the source (upper pannel, black diamonds), the first/second/third neighbors (lower pannel, red disks/blue squares, green triangles). The error bars are standard error of the mean for curves sharing the same rank of neighborhood. The solid lines are fits with Eq.~\ref{HexaFit}. On the upper pannel, a control experiment without any nanopores has been overplotted, showing that the source intensity (magenta crosses) remain constant over time.}
  \label{Fig2}
\end{figure}

  The DIB network is imaged with an AxioZoom (Zeiss) macroscope (magnification 22X), in bright field and epifluorescence (excitation and emission wavelength 444 nm and 555 nm, respectively) equipped with a digital CMOS sensitive camera (Orca Fusion, Hamamatsu, 2048$\times$2048 pixels$^2$) and a XY stage. The macroscope is controlled with Micromanager  to image subsequently the different regions corresponding to the different source droplets. Bright field and fluorescence images are acquired every 4 minutes, for a total duration of about 15 hours.  Bright field images are used to locate the centroids of the droplets, their size, and allow to construct a Delaunay tessellation of the network (see Image analysis details in Supplementary Materials). From this tessellation we obtain the adjacency matrix of the network to determine the rank of neighborhood of each droplet to a given fluorescent source. The printing and compaction process yields a rather monodisperse distribution of droplet sizes (\textit{see} Fig.~\ref{Fig1B}) and a DIB network well approximated by a hexagonal lattice (coordination number $Z\approx$ 6, \textit{see} Fig.~\ref{Fig1B}, inset). Fluorescence images are used to quantify the diffusion of Calcein in the network over time (see Fig.~\ref{Fig2}A-C). To compare the fluorescent intensity $I(x_i,t)$ with the occupation probability $p(x_i,t)$ of a fluorescent molecule at a given lattice site $x_i$, we normalize the intensity as follow :
\begin{equation}
 p(x_i,t) = \frac{I(x_i,t)-I_{bk}}{I(x_S,t=0)} 	
\end{equation} 
\noindent where $ I_{bk}$ is the background intensity taken in a droplet of same rank at time $t=0$ and $x_S$ is the source position (\textit{see} details in the Supplementary Materials).

\section{Results and Modeling}
\subsection{Experimental Results}
\begin{figure*}[htbp]
\centering
  \includegraphics[width=\textwidth]{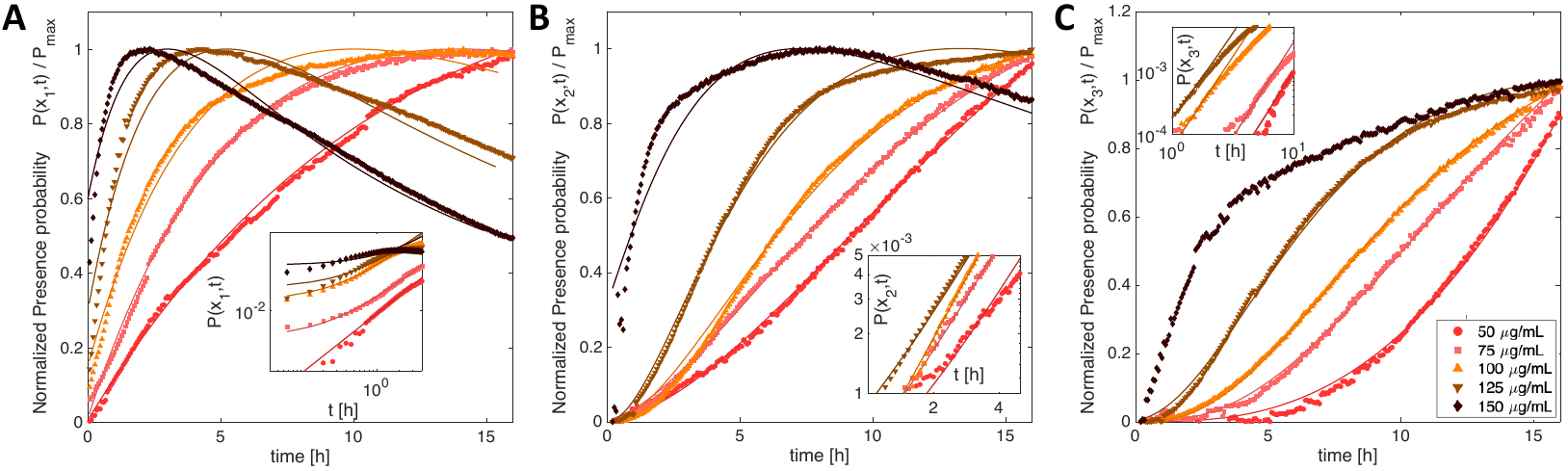}
  \caption{Selection of normalized probabilities $P(t)/P_{max}$ curves for first (A) second (B) and (C) third neighbors. Different colors denote different $\alpha$HL monomer concentrations $c_m$ (see legend on panel C). Solid lines are the best fits using Eq(\ref{HexaFit}). Insets: log-log plot of the presence probability $P(t)$. Solid lines are polynomial fits at short times, using respectively (A) $P\sim \lambda (t-t_0)$ , (B) $P\sim (\lambda (t-t_0))^2$ and (C) $P\sim (\lambda (t-t_0))^3$.}
  \label{Fig3}
\end{figure*}

 We plot on Fig.~\ref{Fig2}A-C fluorescence image of the DIB network with a $\alpha$HL monomer concentration $c_m$ = 125 $\mu$g/mL, taken at increasing times, from top to bottom. Over the course of 15 hours, Calcein has diffused through the network from the source droplet. The corresponding averaged normalized intensity curves of the source droplet, first, second and third neighbors for this experiment are shown on Fig.\ref{Fig2}C and compared to a control experiment ($c_m$ = 0). As the DIBs contain nanopores, the fluorescent molecules initially in the source are transported through the network, whereas for the control experiment, the source intensity remains constant over time. \\
 We plot on Figs.~\ref{Fig3}A-D a selection of normalized intensity curves for different nanopore concentrations, from 50 to 150 $\mu$g/mL, for each rank of neighborhood (source, 1$^\mathrm{st}$ to 3$^{\mathrm{rd}}$ neighbors). The diffusion timescale strongly depends on $c_m$. \\
 \subsection{Theoretical modeling}
 The transport of Calcein in the network is modelled by a Continuous Time Random Walk (CTRW) \cite{hughes1996random}. A given calcein molecule waits a random time to ``jump'' to the next droplet. This random waiting time $\tau$ is qualitatively  the time required for the molecule to reach a nanopore in the membrane, as it will be discussed in more detail later on. In our model, we use an exponential distribution of waiting time, $g(\tau)=\lambda \mathrm{e}^{-\lambda t}$, where $\lambda = 1/<\tau>$ is the inverse of the mean waiting time. Most of the calculation follows the derivation found in \cite{hughes1996random} that we here only briefly recapitulate. Detailed derivations can be found in the Supplementary Materials. \\
 To start, let us consider the case of a \emph{discrete time} random walk. The probability to be at the position $\vec{l}$, at step $n$, starting from the position $0$, is denoted $ P_{n}(\vec{l} \mid \overrightarrow{0})$. At step $n+1$, this probability writes 
\begin{equation}
   P_{n+1}(\vec{l} \mid \overrightarrow{0})=\sum_{l^{\prime}} P_n(\vec{l^{\prime}} \mid \overrightarrow{0}) p(\vec{l}-\vec{l}^{\prime})
\end{equation}
where $p(\vec{l}-\vec{l}^{\prime})$ is the probability to make a jump of $\vec{l}-\vec{l}^{\prime}$ in a single step. \\ 
Seen as a convolution product, the latter  equation writes, in Fourier space,
\begin{equation}
\tilde{P}_{n+1}(\vec{k})=\tilde{P}_n(k) \times \Lambda(\vec{k})
   \label{fourier1}
\end{equation}
with $\Lambda(\vec{k})$ the structure factor of the lattice :
\begin{equation}
\Lambda(\vec{k})=\sum_{l}p(l) e^{-i \vec{l} .\vec{k}}  
\end{equation}
By iteration we get $\tilde{P}_{n}(\vec{k})= {[\Lambda(\vec{k})]}^n$ and taking the inverse (spatial) Fourier transform and the direct (temporal) \emph{discrete} Laplace transform (defined as $\hat{f}(l,\xi)= \sum_{n=0}^{\infty}{f}_{n}(l) \xi^n$)  one finds, in dimension $d$
\begin{equation}
 P(\vec{l}, \xi)=\int \frac{d^d k}{(2 \pi)^d} \frac{e^{-i \vec{k} \vec{l}}}{1-\xi\Lambda(\vec{k})}
  \label{inverselaplaceft}
\end{equation}
The relevant case of a continuous time random walk can be deduced from the discrete time case by use of the \emph{Montrol-Weiss} theorem \cite{montroll1965random,hughes1996random,shlesinger2017origins}, yielding for the \emph{continuous} Laplace transform of the occupancy probability $\hat{p}(\vec{l},s)= \int_{0}^{\infty}  \ p(\vec{l},t)e^{-st}$
\begin{equation*}
\hat{p}(\vec{l},s) = \frac{1-\hat{\Psi}(s)}{s} P\left(l,\hat{\Psi}(s)\right)
\end{equation*}
\noindent where $\hat{\Psi}(s)$ is the Laplace transform of the waiting time probability distribution $g$. 
One finally gets
\begin{equation}
  p(\vec{l},t) =  \frac{1}{4\pi^2} \int_{0}^{\infty} ds \ e^{st} \int  d^2k \ e^{i \vec{k}.\vec{l} }e^{-\left(\lambda\left(1-\Lambda(k)\right)t\right)} 
   \label{Master}
\end{equation}
To fully compute the occupancy probability, one needs an expression for the structure factor $\Lambda(\vec{k})$, which depends on the topology of the lattice. Choosing a square lattice in 2D allows one for a full analytical solution (see Supplementary Materials) for the density of probability to be at any site. Experimentally, however, the topology of the network is not squared, but closer to an hexagonal lattice (Fig.~2, inset). In that case, there is no fully analytical result and integrals have to be evaluated numerically. In hexagonal lattice, one can show easily (see Supplementary Materials) that the structure factor writes $\Lambda(k_1,k_2)=\frac{1}{3} (cos(k_1)+cos(k_2)+ cos(k_1+k_2))$, where $k_1$ and $k_2$ are reciprocal vectors of two Bravais vectors of the network. The probability density to be at site $\vec{l}=(l_1;l_2)$ writes : 

\begin{widetext}
\begin{gather}
p(\vec{l},t) = \frac{1}{4\pi^2} \iint_{- \pi}^{\pi}  dk_1 dk_2 \cos(k_1l_1+k_2l_2) \ e^{-\lambda_0 t \left(1-\frac{  cos(k_1)+cos(k_2)+cos(k_1+k_2)   }{ 3  }\right)  } 
\label{HexaFit}
\end{gather}
\end{widetext}
\noindent that we evaluate numerically. A Taylor expansion at short times can, however, be performed, yielding $p(\vec{l},t) \sim (\lambda t)^{N_i}$, where $N_i$ is the rank of neighborhood of the site $\vec{l}$ with respect to the source. 
 
 \subsection{Fitting the data}
 To fit the experimental occupancy probability with the hexagonal model prediction, we explored different values of $\lambda$. We also had to incorporate an initial delay time $t_0$ to be able to fit the data. This delay time comes from the uncertainty on the initial moment at which calcein diffusion starts, likely due to a delay in membrane formation as droplets are packed together (in practice we allowed values of $t_0 < 2$ hours. Over the whole experiments/droplets we have on average $t_0$=0.7 hours). We therefore use $p(\vec{l},t-t_0) $ from Eq.~\ref{HexaFit},using $t_0$ and $\lambda$ as fit parameters. In practice, we chose a set of  ($t_0,\lambda$) values and computed the chi squared value $\chi^2$ between $p/p_{max}$ from experimental points and from the theoretical predictions of Eq.~ \ref{HexaFit}. Theoretical curves using the fit parameters which minimize the $\chi^2$ values are plotted with solid lines on Fig.~3 with the experimental data, for a selection of experiments at different $\alpha$HL monomer concentrations $c_m$ . The experimental normalized intensities show good agreement with the model predictions, particularly at lower nanopore concentrations ($c_m < 100,\mu\text{g/mL}$), where the fits are most robust. At higher concentrations, the increased and often anisotropic diffusion of Calcein introduces greater variability in the measurements. Furthermore, the onset of diffusion may already have occurred as the imaging begins at high $c_m$, which can reduce the quality of the fits, especially for first-rank neighbors. Thus, we additionally performed polynomial fits of data at short times ($\lambda t $< 0.1) and used Taylor expansions of Eq.\ref{HexaFit} to extract the values of $\lambda$ (See above and Supplementary Materials). Results of the fits (with or without  time delay $t_0$, or Taylor expansion at short times) allow us to extract $\lambda(\vec{l})$ for different nanopores concentrations, as represented on Fig.~5A. On average we observe a power law increase of $\lambda \sim c^n$ with $n = 1.6 \pm 0.5$, an exponent value which will be discussed further below. This power law increase is observed whatever the rank of neighborhood $N_i$ with the source. Note however that a slight systematic increase of $\lambda$ with $N_i$ is observed (see Inset of Fig.4A), but within experimental error bars. Its origin is likely due to a maturation of nanopores insertion in the membrane: third neighbors droplets membrane are ``older", in term of nanopores insertion, than first neighbors membranes.

\begin{figure}[htbp]
  \includegraphics[width=0.45\textwidth]{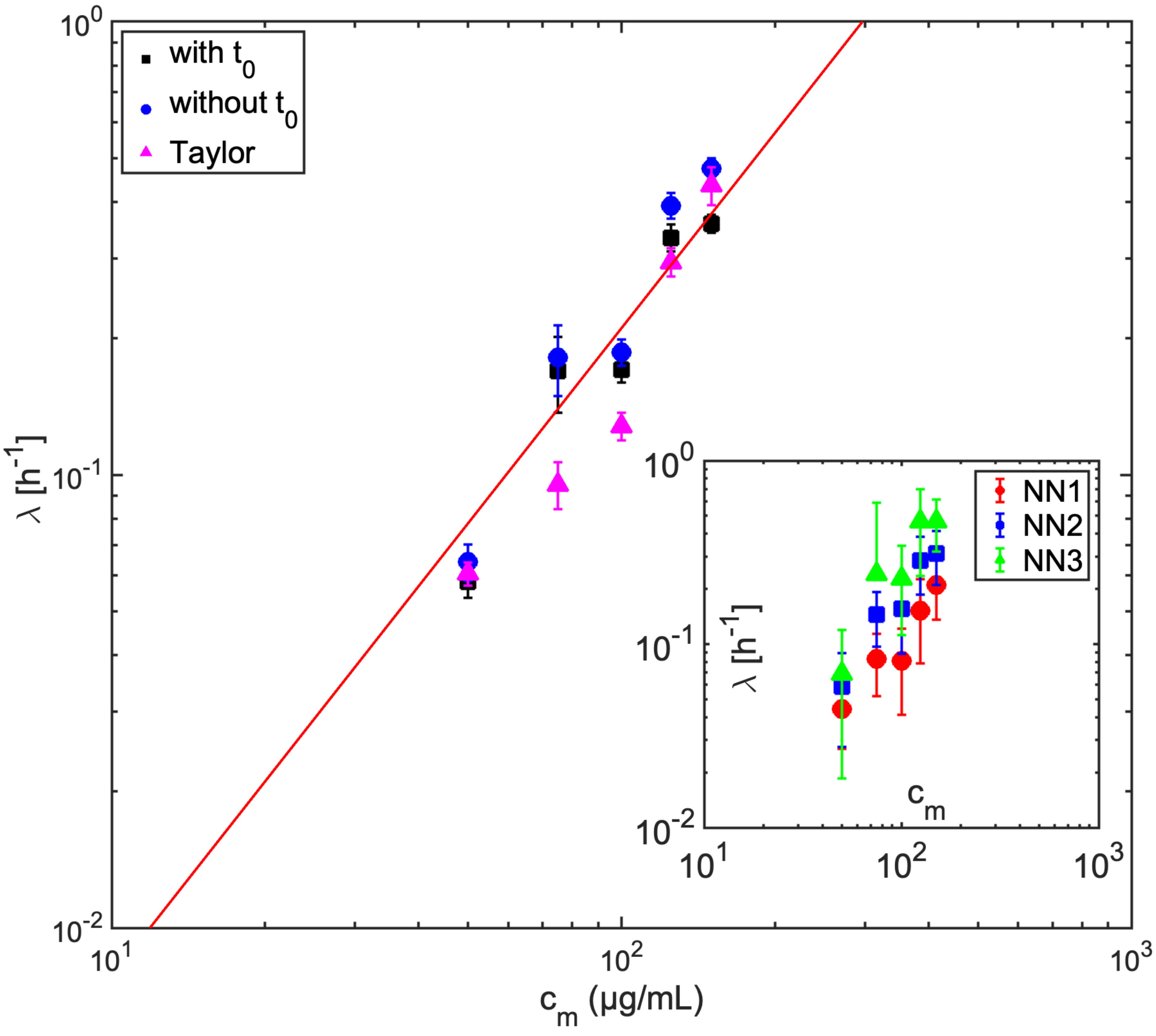}
  \caption{(A) Dependence of the characteristic rate $\lambda$ with the $\alpha$HL monomer concentration. Different symbols are different fit methods to compute $\lambda$. Error bars are standard error on the mean with typically 80 fits per concentration. The line is the best power-law fit of the data $\lambda\sim c_m^n$, with $=1.6\pm 0.2$. Inset: same plot, but differentiating the rank of neighborhood, between first (red), second (blue) and third (green) neighbors. Here, the error bars are standard deviation of data.} 
  \label{fig4}
\end{figure}

\section{Discussion}

One of the advantages of biomimetic approaches lies in their ability to provide quantitative insight into fundamental biological processes, such as transmembrane transport and cell–cell communication, as investigated in this study. Droplet Interface Bilayers (DIBs) functionalized with protein pores such as \(\alpha\)HL, offer a particularly appealing platform. Their simplified and controllable architecture isolates a single transport mechanism—passive diffusion through well-defined pores—allowing quantitative measurements. Changing the concentration of pore-forming monomers in solution, one can tune the average number of pores per membrane, while maintaining structural properties (cell size and shape) and the topology of the mimicked tissue.\\

Diffusion across DIBs functionalized with aHL was implemented in 1D~\cite{valet2019diffusion, Bachler2020} and 2D~\cite{Dupin2018, Elani2014} arrays of DIBs as a way to study not only transport, but also the effect of network topology on coupled chain reactions. 
In many studies, the dynamics of transport itself was not disentangled from that of the chemical reaction, hindering the possibility of quantitative studies and theoretical modelling. 
Conversely, when transport was specifically tackled, two approaches were adopted. On the one hand, a Fick's law can describe the observed dynamics of passage through the porous membrane. In the case of very small compartments, passage through the pore is the kinetic bottleneck and this approach leads to the determination of a rate constant for transport~\cite{Watanabe2016}. However, for this model to universally describe all experimental conditions, it requires the use of an \textit{ad hoc} effective diffusion coefficient and an effective DIB area that is reduced to the cumulated pore area~\cite{Dupin2018}.\\

An alternative description, which we have adopted in this study, consists in modeling molecular transport as a Continuous Time Random Walk (CTRW) on a network of droplets connected via porous membranes. In contrast to the fickian approach, the CTRW framework captures the stochastic nature of both the inter-droplet nanoporosity and the waiting times associated with translocation events. It is well-suited for spatially extended systems, where transport involves not just droplet-droplet permeation but also diffusion between distant compartments. In this work, we extended the 1D transport model developed in \cite{valet2019diffusion} to two-dimensional hexagonal lattices, and to any rank of neighborhood. The approach naturally generalizes to other lattice topologies and can be readily applied to three-dimensional configurations. This makes our model particularly versatile and well suited to study transport phenomena in biological systems.\\

A central observation from our measurements is that the inverse characteristic diffusion time, denoted by \(\lambda\), grows nonlinearly with the concentration of \(\alpha\)-HL monomers in solution. Specifically, we find a power-law dependence of the form \(\lambda \sim c_m^n\), with an exponent \(n \approx 1.6\). This scaling is notably weaker than what we previously reported in 1D DIB arrays \cite{valet2019diffusion}, where \(n\) values close to 3 were obtained. This discrepancy calls for a careful re-examination of the underlying physical picture.
In our previous work on 1D DIBs, we interpreted the observed scaling using a first-passage framework. We postulated that the transport dynamics are governed by the mean waiting time \(\tau\) it takes for a fluorescent molecule to reach a nanopore on the membrane interface. In this framework, a molecule inside a spherical droplet of radius \(R\) must explore the volume of the droplet to find a pore of size \(a\) located on the membrane surface~\cite{chevalier2010first}. For a single pore, this search time scales as \(\tau \sim \frac{R^3}{D a}\), where \(D\) is the molecular diffusion coefficient in the bulk. If \(N\) independent pores are uniformly distributed, the mean waiting time becomes \(\tau \sim \frac{R^3}{N D a}\), reflecting the higher likelihood of encountering a pore. Conversely, if the pores are clustered into a single mesoscopic target of size \(\sqrt{N} a\), the estimate becomes \(\tau \sim \frac{R^3}{D\sqrt{N} a}\). 
To connect this with the experimentally accessible variable—monomer concentration—we invoked a Langmuir-type adsorption model for pore assembly. The formation of a functional \(\alpha\)-HL pore requires the heptamerization of seven monomers, which yields $\langle N \rangle \sim c_m^7$ at equilibrium. Plugging this into the first-passage time estimates, one expects: $\tau \sim c_m^{-7}${(for independent pores)} and $\tau \sim c_m^{-7/2}$  for clustered pores. Both predictions yield exponents significantly larger than what we observe in our 2D DIB network, where \(\tau \sim c_m^{-1.6}\). This substantial mismatch suggests that this simple adsorption/heptamerization model does not fully capture the molecular processes at play in our system. \\
Structural studies indeed suggest other scenarii. Atomic Force Microscopy (AFM) imaging of \(\alpha\)HL in lipid bilayers\cite{czajkowsky1998staphylococcal} reveals the presence of incomplete oligomeric structures, including hexamers, pentamers, or unstable intermediates. These sub-heptameric assemblies may insert into the membrane but fail to form functional channels. Moreover, electrophysiological recordings indicate considerable variability in the conductance of formed pores, particularly under non-optimal conditions such as low temperature or complex lipid compositions\cite{menestrina1994pore} . This heterogeneity suggests that only a subset of oligomers contributes effectively to molecular transport.\\
The heptamerization mechanism in biological membranes may be sensitive to physicochemical parameters—such as curvature, membrane tension, or lipid composition—that are not yet accounted for in existing models. Beyond the distinction between 1D and 2D systems, a notable difference between the DIBs used in~\cite{valet2019diffusion}  and those employed in the present study lies in their lipid composition: the former consisted exclusively of DPhPC bilayers, whereas our system relies on a four-lipid mixture. Lipid composition could influence pore heptamerization within the membrane, and thereby modulate the dependence of the characteristic diffusion time on monomer concentration. Previous studies have highlighted that variations in lipid composition can significantly alter protein adsorption, conformational stability, and insertion efficiency within membranes~\cite{dominguez2016impact,hickey2011lipid,lee2014lipid}. Furthermore, the local curvature of the membrane and its lateral tension are known to modulate the free energy landscape of protein–lipid interactions, thereby potentially affecting both the kinetics and thermodynamics of oligomerization processes~\cite{zimmerberg2006proteins,golani2019membrane}.\\
Based on these insights, we may postulate that multiple oligomerization pathways are at play. For instance, if one combines a heptamerization reaction with a pentamerization one, a simple kinetic model can reduce the dependence of heptamer concentration on the monomer concentration (see Supplementary Material). As a result, the actual number of effective transport channels, $N_{\text{eff}}$, can become significantly smaller than predicted by a simple heptamerization model, and $N_{\text{eff}}$ increases sub-linearly with monomer concentration: $N_{\text{eff}} \sim c_m^\beta$. This suggests that the nonlinear scaling we observe arises from the complex and inefficient formation of pore units within the membrane, rather than from the geometric distribution of the pores.\\
Altogether, these results call for systematic studies to quantify how nanopore insertion in DIBs is affected by lipid composition. They also suggest that altering the way proteins insert into cellular membranes can have strong effects on transport at the multicellular scale, highlighting the need to revisit transport modeling in tissues with these new considerations.

\section{Conclusion}
Our work provides experimental methods to construct quasi-hexagonal networks of droplet interface bilayers (DIBs) whose lipid membranes are decorated with transmembrane nanopores. Using fluorescence imaging, we quantified the diffusion of calcein molecules through the network over time at varying concentrations of pore-forming monomers. To interpret these dynamics, we developed a Continuous Time Random Walk model on hexagonal lattices, which successfully reproduces the spatiotemporal evolution of calcein concentration profiles at any lattice distance from the source. We find that the characteristic diffusion time decreases as a power law with the concentration of pore-forming monomers, with an exponent that cannot be explained solely by pore clustering arguments. Our results suggest instead that the lipid membrane composition may play a key role in controlling the adsorption efficiency of pores, thereby modulating the effective diffusion kinetics within the network.

\section*{Acknowledgements}
The authors acknowledge Raphaël Voituriez and Léo Regnier for their help in the theoretical modeling of first passage time statistics.
\bibliographystyle{rsc} 
\bibliography{rsc} 

\end{document}